\documentclass{article}

\usepackage{arxiv}

\usepackage[utf8]{inputenc} 
\usepackage[T1]{fontenc}    
\usepackage{hyperref}       
\usepackage{url}            
\usepackage{booktabs}       
\usepackage{amsfonts}       
\usepackage{nicefrac}       
\usepackage{microtype}      
\usepackage{lipsum}		
\usepackage{graphicx}
\usepackage{natbib}
\usepackage{doi}
\usepackage{siunitx}
\usepackage{svg}

\usepackage{tablefootnote}
\usepackage{subfig}

\usepackage{xargs}                      
\usepackage{todonotes}

\title{A smile is all you need: \\ Predicting limiting activity coefficients from SMILES \\ with natural language processing }

\author{ \href{https://orcid.org/0000-0001-5764-9757}{\includegraphics[scale=0.06]{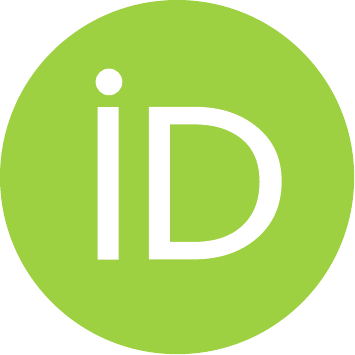}\hspace{1mm}Benedikt A.~Winter}\\
	Energy and Process System Engineering\\
	ETH Zürich\\
	Tannenstrasse 3, 8092 \\
	Zürich, Switzerland \\
	\texttt{bewinter@ethz.ch} \\
	\And
	{\hspace{1mm}Clemens S.~Winter} \\
	OpenAI\\
	3180 18TH St, CA 94110\\
	San Francisco, USA\\
	\texttt{clemenswinter1@gmail.com} \\
	\And 
	\href{https://orcid.org/0000-0001-8013-5439}{\includegraphics[scale=0.06]{orcid.pdf}\hspace{1mm}Johannes ~Schilling}\\
	Energy and Process System Engineering\\
	ETH Zürich\\
	Tannenstrasse 3, 8092 \\
	Zürich, Switzerland \\
	\texttt{jschilling@ethz.ch} \\
	\And 
	\href{https://orcid.org/ 0000-0002-3831-0691}{\includegraphics[scale=0.06]{orcid.pdf}\hspace{1mm}Andr\'e ~Bardow}\thanks{Corresponding author} \\
	Energy and Process System Engineering\\
	ETH Zürich\\
	Tannenstrasse 3, 8092 \\
	Zürich, Switzerland \\
	\texttt{abardow@ethz.ch} \\
}

\renewcommand{\shorttitle}{A SMILE is all you need}

\hypersetup{
pdftitle={A SMILE is all you need Winter et al. 2022},
pdfsubject={property prediction},
pdfauthor={Benedikt~Winter},
pdfkeywords={thermodynamic properties, machine learning, activity coefficients, SMILES},
}

\begin{document}
\maketitle

\begin{abstract}
Knowledge of mixtures' phase equilibria is crucial in nature and technical chemistry. Phase equilibria calculations of mixtures require activity coefficients. However, experimental data on activity coefficients is often limited due to high cost of experiments. For an accurate and efficient prediction of activity coefficients, machine learning approaches have been recently developed. However, current machine learning approaches still extrapolate poorly for activity coefficients of unknown molecules. In this work, we introduce the SMILES-to-Properties-Transformer (SPT), a natural language processing network to predict binary limiting activity coefficients from SMILES codes. To overcome the limitations of available experimental data, we initially train our network on a large dataset of synthetic data sampled from COSMO-RS (10 Million data points) and then fine-tune the model on experimental data (\num{20870} data points). This training strategy enables SPT to accurately predict limiting activity coefficients even for unknown molecules, cutting the mean prediction error in half compared to state-of-the-art models for activity coefficient predictions such as COSMO-RS, UNIFAC\textsubscript{Dortmund}, and improving on recent machine learning approaches. 
\end{abstract}

\keywords{thermodynamic properties \and machine learning \and activity coefficients \and SMILES}

\section{Introduction} \label{sec:intro}
With over $500\,000$ molecules registered even in the CAS Common Chemicals database \citep{CAS.06.02.2022}, the chemical design space of molecules is substantially larger than our capacity to measure their thermodynamic property data. This gap further increases when considering that properties usually depend on temperature and pressure, and even more for mixtures due to combinatorics and dependency on mixture composition. Binary activity coefficients are of particular interest in chemical engineering, as activity coefficients govern the phase equilibria in distillation and extraction, the key separations of many chemical processes. However, even large property databases, such as the Dortmund Datenbank (DDB), only hold experimental data for the activity coefficients of \num{31000} binary systems, a tiny fraction of all possible molecular combinations\citep{DortmundDatenbank.2022}.

To overcome the inherent lack of experimental data, predictive thermodynamic property models have been developed over recent decades for many molecular properties, e.g., COSMO-RS \citep{Klamt.1995}, COSMO-SAC \citep{Lin.2002}, SAFT-$\gamma$ Mie \citep{Lafitte.2013}, and UNIFAC \citep{Fredenslund.1975}. These models can predict thermodynamic properties with increasing accuracy and are therefore particularly beneficial for molecule mixtures with missing experimental data. However, despite the vital advantages of predictive thermodynamic models, these models come with shortcomings. E.g., calculating the surface charges of molecules for COSMO models is time-consuming, whilst UNIFAC is limited to known functional groups parametrized to experimental data. Moreover, these physically-based  predictive models are still less accurate than experiments\citep{Brouwer.2019}. 

Computationally efficient alternatives to physically-based predictive models are data-driven models using machine learning. Machine learning is currently a rising topic in chemical engineering, as summarized in multiple recent reviews \citep{Alshehri.2022,Haghighatlari.2019,Dobbelaere.2021} that identify challenges in many areas such as optimal decision making, introduction and enforcing of physics, information and knowledge representation, and safety and trust \citep{Schweidtmann.2021}. The application of machine learning has also already led to recent advances in thermodynamic property prediction. \citet{Alshehri.2021} developed a data-driven model to predict 25 pure component properties based on a Gaussian process. The developed model surpasses classical group contribution models in accuracy. \citet{Chen.2021} use a transformer-convolutional model to predict the sigma profiles of pure components with high accuracy. 

To predict activity coefficients, matrix completion methods have been recently proposed that represent the limiting activity coefficient of binary mixtures as a matrix. In matrix completion methods, all mixtures are sorted into a solvent-by-solute matrix. Known mixtures are used to learn embeddings for each solvent/solute, which then can be used to fill the matrix by interpolating towards unknown combinations. \citet{Jirasek.2020} proposed a matrix completion method to predict limiting activity coefficients of binary mixtures at 298.15 K that exceeded the accuracy achieved by UNIFAC.  Recently, \cite{Damay.2021} extended the method of \citet{Jirasek.2020} to capture temperature dependencies. The proposed model has a higher accuracy for the temperature-dependent prediction of limiting activity coefficients than UNIFAC. \cite{Chen.2021b} developed an approach to extend the UNIFAC-Il model \citep{Nebig.2010} for predicting limiting activity coefficients in ionic liquids by combining matrix completion with convolutional networks. These proposed approaches exceed the accuracy of the widely employed UNIFAC model in predicting limiting activity coefficients. Moreover, matrix completion approaches do not require any characterization of the molecules to train the model and predict thermodynamic properties, as the model solely learns from the correlations within the matrix. However, their lack of molecular characterization prevents matrix completion methods from extrapolating beyond the space of molecules available for training. Recently, \cite{SanchezMedina.2022} developed a graph neural network to predict limiting activity coefficients at constant temperature. In principle, this graph neural network is capable of extrapolating to unknown solvents and solutes, but the extrapolatory capabilities of the network were not tested. Thus, it is still unclear how well machine learning methods can extrapolate out of the realm of training data onto unknown solutes and solvents. 

Here, we present the SMILES-to-Property-Transformer (SPT), a data-driven model with high accuracy for interpolation and extrapolation that can predict temperature-dependent limiting activity coefficients from nearly arbitrary SMILES, based on natural language processing and a transformer architecture \citep{Vaswani.2017}. Due to their ability to learn structural relationships, transformer models have recently shown to be successful in predicting pure component properties of various molecules and pharmaceuticals \cite{Rong.18.06.2020,Skinnider.2021}. However, transformer models require large amounts of training data, which is typically unavailable for thermodynamic properties from experiments. To overcome the lack of experimental training data, we propose a two-step approach: First, the model is trained on a large amount of synthetic data from a physically-based predictive model for limiting activity coefficients to convey the grammar of SMILES and the underlying physics of activity coefficients to the model. Second, the pretrained model is fine-tuned using available experimental data to improve accuracy and reduce systematic errors of physically-based predictive model. 
We compare the SPT model to state-of-the-art predictive thermodynamic models and ML approaches and demonstrate its high accuracy for predicting temperature-dependent limiting activity coefficients of unknown molecules after fine-tuning.

\section{Transformer-based method for thermodynamic property prediction} \label{sec:arc}

The SPT model predicts temperature-dependent limiting activity coefficients of binary mixtures from the SMILES codes of the mixture components. For this purpose, we apply a transformer model. For machine learning, two major characteristics are vital for success: the model's architecture and the training data. We first describe our model architecture (Section~\ref{sec:02_architecture}) and subsequently discuss the datasets used for training and validation of the model (Section~\ref{sec:data}), data augmentation (Section~\ref{sec:aug}) and model parametrization (Section~\ref{sec:hyper}). 

\subsection{Architecture of the SMILES-to-Property-Transformer} \label{sec:02_architecture}

The SPT model is based on the transformer architecture developed by \cite{Vaswani.2017} for natural language processing. Since its conception in 2017, the transformer architecture has proven to be applicable to many tasks beyond natural language processing, such as image generation or classification \citep{Parmar.15.02.2018, Dosovitskiy.22.10.2020}. For molecular property prediction, the transformer model has been successfully applied to predict pure component properties for various pharmaceuticals by \cite{Lim.2021} or generate novel molecules with specific target properties \citep{Kim.2021}. However, to the best of the authors' knowledge, the transformer model has not yet been applied to predict thermodynamic properties of binary mixtures.

\begin{figure}
    \centering
    \includegraphics[width=0.5\textwidth]{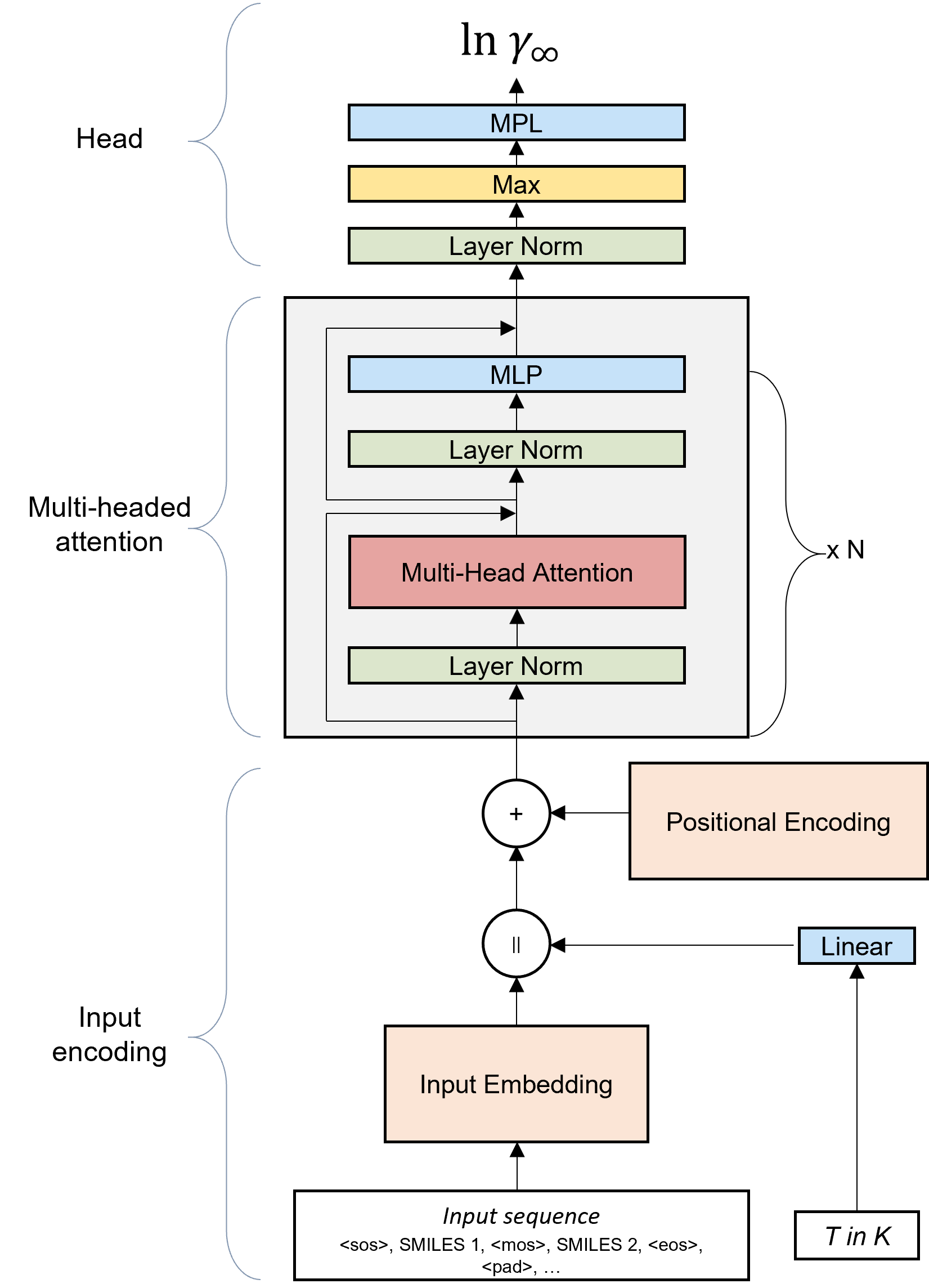}
    \caption{Architecture of SPT to predict limiting binary activity coefficients from SMILES codes. The model takes the input sequence consisting of the SMILES of the solvent and solute and the temperature as input. In the Input encoding section of the model, the information about the entering SMILES, the temperature and the position of tokens are all compiled into a single matrix. The multi-headed attention section performs the main work of the model by transmitting information between different parts of the molecules. The head section reduces the multidimensional output of the model to a single value.}
    \label{fig:network}
\end{figure}

As the backbone of SPT, we adopt the GPT-3 architecture \citep{Brown.28.05.2020} as implemented by \cite{Karpathy.2021} in MinGPT with changes to the input encoding and head section of the model  (Figure~\ref{fig:network}). The GPT-3 architecture shows higher accuracy than, e.g., the transformer implementation of PyTorch \citep{Pytorch.2021}, most likely due to the use of a Post-LN Transformer instead of a Pre-LN Transformer \citep{Xiong.}. 

\subsubsection{Input encoding} \label{sec:input}

Calculating the temperature-dependent limiting activity coefficients of a solute in a solvent requires information about the structure of both molecules and the temperature. In our model, the molecules are represented by SMILES codes. The Simplified Molecular-Input Line-Entry System code, abbreviated to SMILES, was introduced in 1988 by \cite{Weininger.1988} as a method to represent complex molecules in a single line of text. Since then, the SMILES code has been used in many applications and has developed into one of the standard ways to represent molecules. In SMILES, heavy atoms are represented as their periodic table symbol, e.g., C for carbon, while hydrogen atoms are implicitly assumed, e.g., ethane has the SMILES code CC. For single bonds, atoms are simply chained together, while double or triple bonds are represented by = and \#, respectively. Branching arms of a molecule are contained within brackets, and for rings, numbers are used to show the joining points of a ring. Thus, the molecule 2-methyl phenol can be represented by the following SMILES: Oc1c(C)cccc1. Since SMILES essentially possess a grammar to convey the structure of a molecule in a linear form, they can be understood by natural language processing. Thus, SMILES have shown to be a suitable input for deep learning models that predict molecular properties \citep{Tetko.2020, Wang.2019}.

In the first step of our model, only the molecule's structural information is passed  to the model by constructing an input sequence from the SMILES of the solute and the solvent. Four special characters are used to signal 1) the start of the first molecule,  <SOS>, 2) the middle between both molecules <MOS>, 3) the end of the second molecule <EOS>, and 4) the padding <PAD> to fill the input sequence to a fixed length of $n_\mathrm{seq}$ , e.g.: 

\begin{equation}
\centering
    <\mathrm{SOS}>,\mathrm{SMILES_{solute}},<\mathrm{MOS}>,\mathrm{SMILES_{solvent}},<\mathrm{EOS}>,<\mathrm{PAD}>,...
\end{equation}

Next, the input sequence is tokenized by assigning a number to each unique character of the SMILES code. In general, each token could be longer than a single character per encoding. However, a single character is used in this work for simplicity. Consequently, the vocab contains the following tokens: <SOS>, <MOS>, <EOS>, <PAD>, characters that can be contained in a SMILES code adapted from \cite{Kim.2021} and a special token for water to clearly distinguish between pure water (SMILES "O") and oxygen groups on hydrocarbons. Not all tokens included in the vocab are part of the molecules of our training data. Thus, the embedding of some tokens remains untrained in our final model. Including these untrained tokens makes the model easily expandable for more complex structures in later fine-tuning steps. However, evaluating SMILES that contain untrained tokens leads to unreliable results. The overall vocab and a list of trained and untrained tokens are available in the Supporting Information S1.

After tokenizing the input sequence characterizing the solute and solvent, the input matrix $X$ is constructed from the input sequence and the embedding matrix $E$. The embedding matrix $E$ contains a learned vector of length $d_\mathrm{emb}$ for each token. The input matrix $X$ is constructed by concatenating the embedding vectors belonging to the input tokens resulting in an $n_\mathrm{seq} \times d_\mathrm{emb}$ matrix. Next, temperature information is incorporated into the model by projecting the temperature into the embedding space via a linear layer and concatenating it to the right of the input matrix. Thereby, the matrix size is expanded to $n_\mathrm{seq+1} \times d_\mathrm{emb}$. Positional information is included by adding the learned positional encoding matrix  $D$ of size $n_\mathrm{seq+1} \times d_\mathrm{emb}$ to $X$. Next, the input matrix $X$, which now contains all information about the molecules and temperature, is passed to the transformer block, the heart of the model.

\subsubsection{Transformer block: Multi-headed attention} \label{sec:tras}

In the transformer block, the inputs are normalized via a layer norm and then passed to the multi-headed attention block. On a high-level, multi-headed attention allows the model to move information from one token to another. For molecular property prediction, multiple attention heads enable each head's attention to focus on different features. This attention mechanism can learn complex structures of molecules even when represented as a linear string. On a mathematical level, the output of a single attention head $i$, $Z_i$, is defined as:

\begin{equation}
Z_i = \mathrm{softmax} \left( \frac{Q_iK_i^\mathrm{T}}{\sqrt{d_k}} \right) V_i
\end{equation}

with the query matrix $Q_{i}$, the key matrix $K_{i}$, and the value matrix $V_{i}$, and $d_\mathrm{k} = d_\mathrm{emb} / n_\mathrm{head}$, where $n_\mathrm{head}$ is the number of attention heads.

The query, key, and value matrices are calculated by multiplying the input matrix $X$ with the learned matrices $W_{i}^\mathrm{Q}$, $W_{i}^\mathrm{k}$, and $W_{i}^\mathrm{V}$. The matrices $Q_{i}$, $K_{i}$, and $V_{i}$ have the size $n_\mathrm{seq+1} \times d_\mathrm{k}$. The product of $Q_{i}$ and $K_{i}$ can be interpreted as the relative importance of each token to another token.   This result is normalized by the square root of the key dimension $d_\mathrm{k}$ and passed to a softmax function returning the attention from each token to each other token. The value matrix is then multiplied with the attention, leading to the matrix $Z_i$ of size $n_\mathrm{seq+1} \times d_\mathrm{k}$, which contains for each token information of other tokens weighted by their importance.   

The attention operation is repeated for each attention head. The resulting output matrices $Z_i$ of each attention operation are concatenated and projected to the input size of  $n_\mathrm{seq+1} \times d_\mathrm{emb}$ via a linear layer. Finally, the data is passed through a multilayer perceptron (MLP) layer containing a GeLu non-linearity, concluding the transformer block. In the first MLP layer, the size of the model is increased by a factor of four for the embedding dimension; the second linear layer of the MLP projects it back down to the input size. Residual connections connect the in- and output of the attention and MLP block, including their respective layer norms. Multiple transformer blocks can be stacked consecutively to increase the depth of the model. In this work, we use two consecutive transformer blocks. For a more in-depth and visual explanation, the reader is referred to the blog of \cite{Alammar.2018}.

\subsubsection{Head} \label{sec:head}

The output of the transformer blocks needs to be projected to a single value. This projection is performed in the last part of the model, the head. The head first applies a max function along the sequence dimension that reduces the size from $n_\mathrm{seq+1} \times d_\mathrm{emb}$ to $1 \times d_\mathrm{emb}$, followed by one MLP that reduces the size from $1 \times d_\mathrm{emb}$ to $1 \times 1$. The resulting single output value represents the molecular property of interest, i.e., the limiting activity coefficient in our work.

\subsection{Property data for training and validation} \label{sec:data}

While machine learning models have proven to be powerful tools capable of astonishing predictions, their training requires large amounts of data. Such large amounts of data are typically unavailable for binary property data. Some success has been achieved using unsupervised learning on SMILES translation tasks to pretrain the model \citep{Honda.12.11.2019}. In this work, we propose using synthetic data for the molecular property of interest for pretraining. Subsequently, we use experimental data for the fine-tuning of the model. The definition of the training and validation sets for pretraining and fine-tuning is shown in Figure~\ref{fig:data} and explained in the following section.

\subsubsection{Synthetic data for pretraining:} \label{sec:syn}

\begin{figure}
    \centering
    \includegraphics[width=\textwidth]{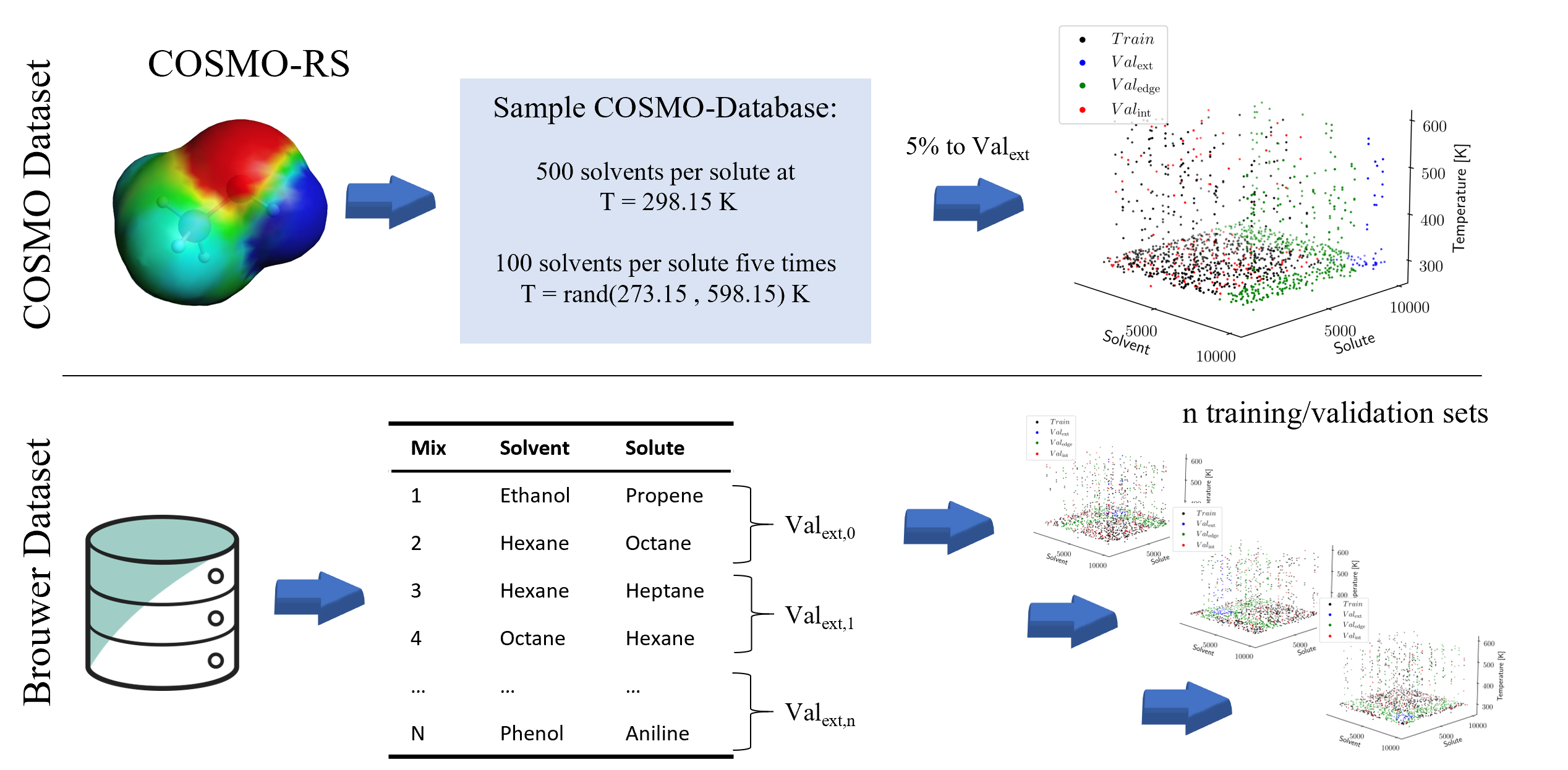}
    \caption{Creation of the synthetic dataset from COSMO used for pretraining and sampling procedure for the experimental data from Brouwer used for fine-tuning. For the COSMO dataset, \SI{5}{\percent} of the solvents and solutes are removed from the training set constructing $\mathrm{Val}_\mathrm{ext}$, for which neither solvent nor solute is known, and $\mathrm{Val}_\mathrm{edge}$ for which either solvent or solute is known. Furthermore, \SI{5}{\percent} of the remaining training data is sampled at random and moved into the training set $\mathrm{Val}_\mathrm{int}$. Due to the smaller size of the Brouwer dataset, n-fold cross-validation is used. $N$ mixtures are selected and moved into $\mathrm{Val}_{\mathrm{ext,}i}$, resulting in $n = 1000$ validation sets. The remaining mixtures are sorted into  $\mathrm{Val}_{\mathrm{edge,}i}$, and $\mathrm{Train}_i$ depending on the occurrence of their constitutes in $\mathrm{Val}_{\mathrm{ext,}i}$. Finally, \SI{5}{\percent} of mixtures are removed from $\mathrm{Train}_i$ to $\mathrm{Val}_{\mathrm{int,}i}$, and the set of all mixtures is reassessed.}
    \label{fig:data}
\end{figure}

For the pretraining of our model, we generate a large amount of synthetic data using the established thermodynamic model COSMO-RS \citep{Klamt.1995}. The advantage of COSMO-based models is that they can predict activity coefficients for arbitrary molecules from the molecular structure and are not limited to specific functional groups such as UNIFAC (\cite{Fredenslund.1975}). Thus, training data can be generated from a more diverse set of molecules, increasing the machine learning model's ability to extrapolate. Furthermore, an extensive database and infrastructure to sample COSMO-RS are available from our previous work \citep{Scheffczyk.2016}.

To generate the synthetic data, we use the COSMObase2020 database. This database contains around $10\,000$ molecules resulting in more than 100 million possible binary combinations for solutes and solvents. Calculating activity coefficients for all combinations is computationally intractable. Thus, for each of the $10\,000$ solutes, 500 random solvents are sampled at a temperature of $T = \SI{298.15}{\kelvin}$, resulting in around 5 million solvent/solute combinations. Furthermore, 100 of the 500 random solvents per solute are sampled at five random temperatures between \SI{273.15}{\kelvin} and \SI{598.15}{\kelvin} to provide temperature-dependent data. In total, around 10 million data points are sampled for pretraining, referred to as the COSMO dataset. We use the TZVDP-FINE parametrization and a maximum of 3 conformers for calculating the limiting activity coefficient of each data point.

To validate the performance of our machine learning model during the pretraining, the COSMO dataset is split into three validation sets. For this purpose, \SI{5}{\percent} of the solvents and solutes are initially removed from the training set (see Figure~\ref{fig:data}). Crucially, preliminary tests showed that water cannot be entirely removed from the training set to ensure an accurate prediction for this notable molecule. Removing solvents and solutes from the training set enables the creation of two validation sets: first, a validation set containing the cross-section of the excluded solvent and solutes, where the training data contains neither the solvent nor the solute. This validation set tests the extrapolation accuracy of the model for entirely unknown solute/solvent combinations and is referred to as Val$_\mathrm{ext}$. Second, a validation set is created where either solvent or solute is contained in the training set, but not both. This validation set tests the extrapolation capability of the model for one unknown molecule while the other one is known. Since the validation set tests the edge of known structures, we call it Val$_\mathrm{edge}$. Lastly, an additional \SI{5}{\percent} of the remaining solute-solvent combinations are randomly removed from the training set. If a solute-solvent combination exists for more than one temperature, the combination is removed for all temperatures. The resulting third validation set, so-called Val$_\mathrm{int}$, tests the interpolation capabilities of the model when solvent and solute are known in other combinations but not in precisely this combination. This validation set is most comparable to the matrix completion approaches discussed earlier, where both mixture components have to be known. 

\subsubsection{Experimental data for fine-tuning:}
\label{sec:data_exp}

In the second step, the model pretrained on the COSMO dataset is fine-tuned to experimental data. To increase the reproducibility and accessibility of our model, we solely use publicly available data on limiting activity coefficients. Furthermore, using open-source experimental data enables an open benchmark to compare other methods.

To our knowledge, the largest publicly available dataset on limiting activity coefficients was published recently by \cite{Brouwer.2021}. This dataset contains \num{77173} limiting activity coefficients for various solute/solvent combinations and temperatures gathered from the literature. However, from the $77\,173$ data points, around $52\,000$ data points use ionic liquids or deep eutectic solvents as solvents and are thus excluded. Additionally, we excluded impure substances such as sunflower oil, solvents with specific phase orientations (nematic phase, isotropic phase), and uranium complexes. For 10 solvents/solutes, no SMILES code could be identified. Furthermore, some errors in the data by \cite{Brouwer.2021} were corrected, such as wrong exponents, $\ln \gamma^\infty$ instead of $\gamma^\infty$, misclassification, or data entered in the wrong row. A list of all changes and an updated data table can be found in the Supporting Information S2. Overall, $20\,870$ suitable data points are identified and used for the fine-tuning of our model. The resulting data set for the fine-tuning contains \num{349} solvents and \num{373} solutes in \num{6416} unique combinations in a temperature range from \SI{250}{\kelvin} to \SI{555.6}{\kelvin}. The distribution of the data in $\ln \gamma^\infty$ and $T$ is shown in the Supporting Information S3. In the following, the dataset is referred to as the Brouwer dataset.

To test the performance of the fine-tuning, again, three validation sets are defined as for the pretraining. Due to the much smaller amount of data available from experiments, n-fold cross-validation is used to determine the accuracy of the network. Due to the small sample size of a single validation set, this approach would be expected to have a high variance (Figure \ref{fig:data}). To construct the training and validation sets, all solute/solvent combinations without water are split into 1000 subsets, each constructing one Val$_{\mathrm{ext},i}$. The solute/solvent combinations not part of Val$_{\mathrm{ext},i}$ are assigned either to the edge validation set Val$_{\mathrm{edge},i}$, or the training set Train$_i$ depending on whether one or none of the two components are part of the Val$_{\mathrm{ext},i}$. Subsequently, \SI{5}{\percent} of the training set Train$_i$ is randomly sampled to yield the validation set Val$_{\mathrm{int},i}$ used to test the interpolation capability. Finally, all data points are reassessed to determine whether they have to be moved to another validation set due to the removal of Val$_{\mathrm{int},i}$ from Train$_i$. 

Solvent-solute combinations with water are excluded from Val$_{\mathrm{ext},i}$ for two main reasons: First, the unique nature of water makes it challenging to extrapolate water properties when only organic compounds are known within the training set. Second, we believe that applications are rare where the limiting activity coefficient of the unknown and unmeasured molecule water must be predicted. While water is excluded from the validation set Val$_{\mathrm{ext}}$, the validation set Val$_{\mathrm{edge}}$ still contains combinations with water as a known solvent and an unknown organic solute, which we envisage as likely use-cases. Results for the validation set Val$_{\mathrm{int}}$ and Val$_{\mathrm{edge}}$, including only combinations with water, are available in the Supporting Information S6.

Due to the varying amount of data points for each solute/solvent combination, the size of the training sets varies. The sizes range between $15\,000$-$19\,270$ for Train, $6$-$69$ for Val$_{\mathrm{ext}}$, $640$-$5\,000$ for Val$_{\mathrm{edge}}$, and $640$-$1\,200$ for Val$_{\mathrm{int}}$.

\subsection{Data augmentation} \label{sec:aug}

We increase the variety of the data provided to the model by generating up to 9 equivalent SMILES for each input molecule using the tool of \cite{Bjerrum.21.03.2017}. During training, one of the resulting 10 SMILES is randomly selected to construct the input sequence. During validation, the initially assigned SMILES are used.

\subsection{Training and hyperparameter tuning} \label{sec:hyper}

Identifying good hyperparameters is vital   for the performance of machine learning models. We select hyperparameters by conducting a manual scan on the COSMO dataset, considering embedding size, number of heads, number of attention layers, dropout, batch size, and learning rate. The loss function is fixed to mean-squared-error (MSE) loss. The Adam optimizer and cosine annealing with linear warmup is used as a learning rate schedule with a warmup time of 5 epochs. For the hyperparameter tuning, training was stopped after 20 epochs, while the final pretraining ran for 50 epochs. The model is trained in mixed precision with the pytorch autocast function to reduce training time. A detailed hyperparameter table is available in S4.

\section{Results: Predicting limiting activity coefficients} \label{sec:res}

Our machine learning model STP is trained on synthetic and experimental data to predict limiting activity coefficients, as described in Section~\ref{sec:data}. In this section, we first introduce the results of the pretraining to synthetic data (Section~\ref{sec:pre}). Then, we discuss the final results based on fine-tuning to experimental data (Section~\ref{sec:fine}).

\begin{figure}[b]
\centering
    \subfloat[Val$_\mathrm{int}$]{
    \includegraphics[width=0.33\textwidth]{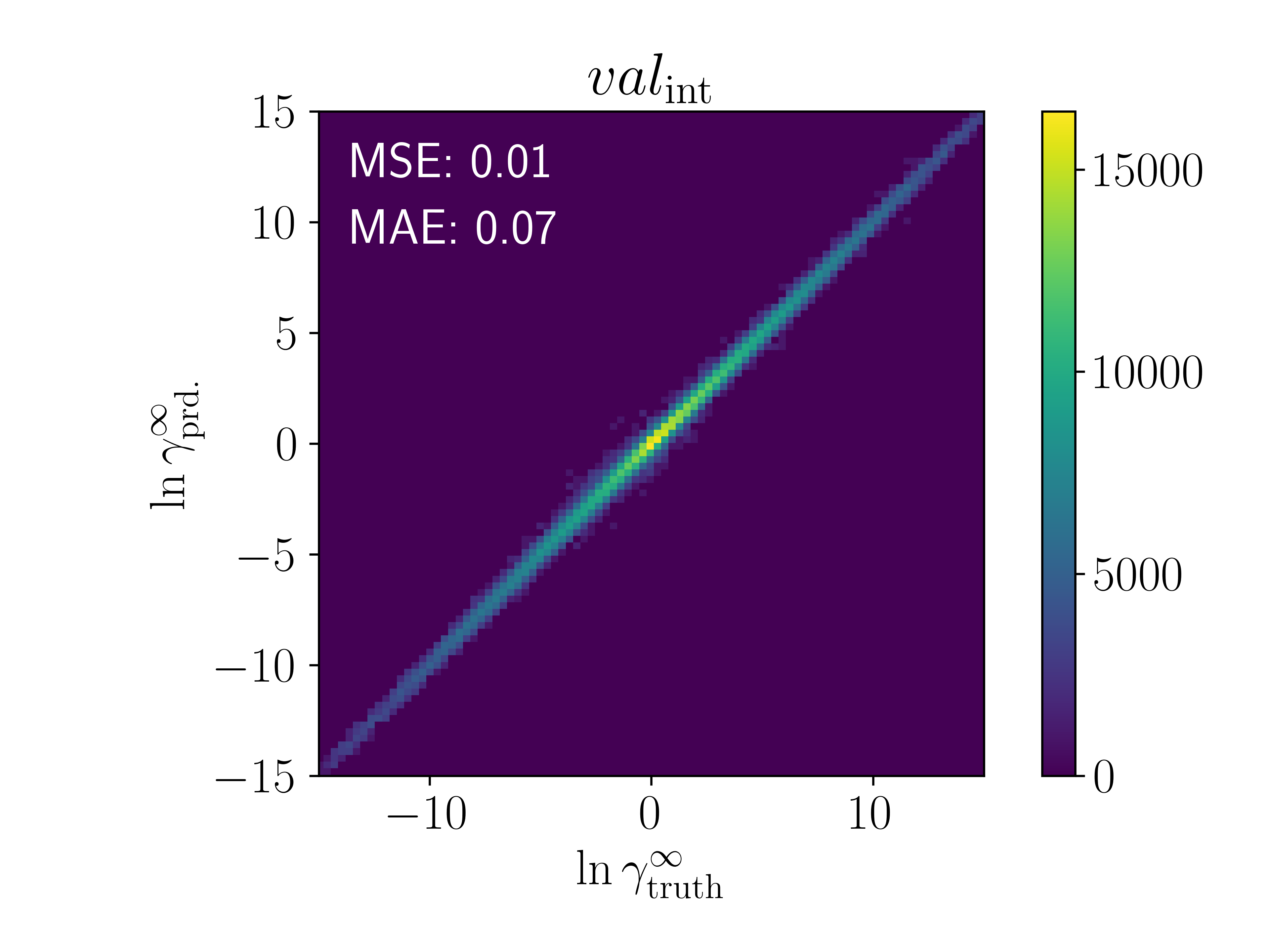}%
    }
    \subfloat[Val$_\mathrm{edge}$]{
    \includegraphics[width=0.33\textwidth]{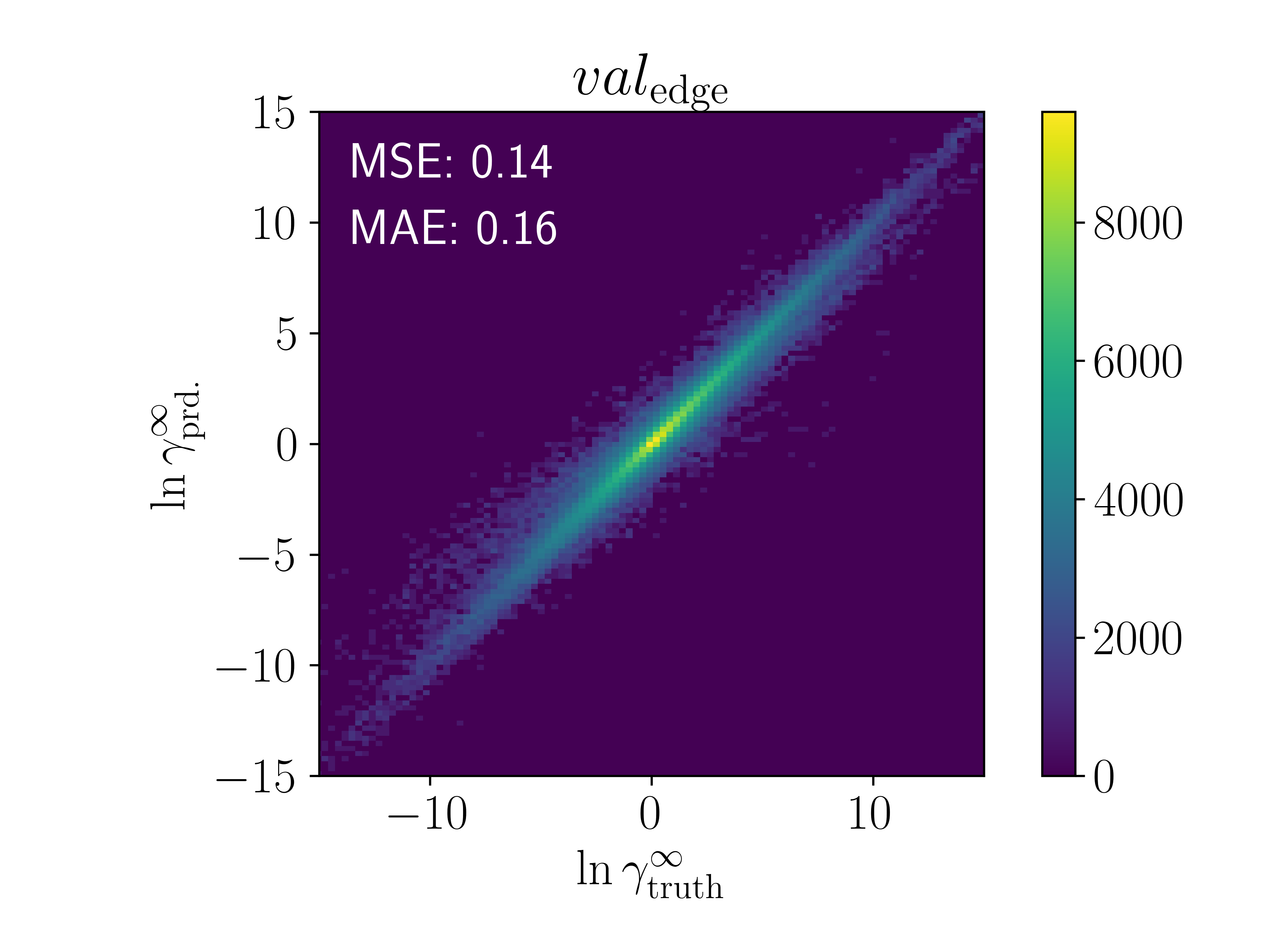}%
    }
    \subfloat[Val$_\mathrm{ext}$]{
    \includegraphics[width=0.33\textwidth]{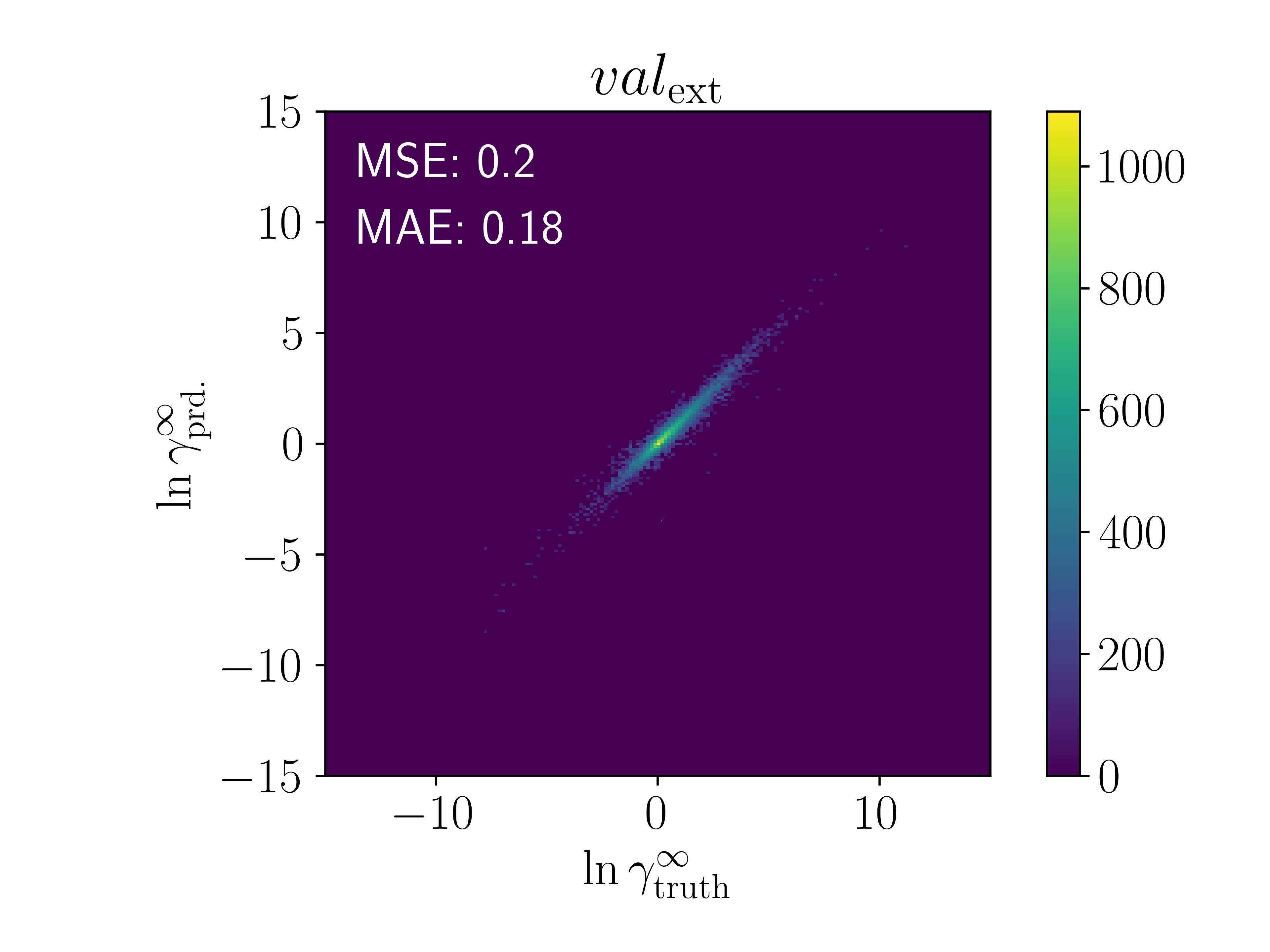}%
    }
    \caption{Heatmap of predicted limiting activity coefficients $\ln \gamma^\infty_\mathrm{pred.}$ vs the validation data $\ln \gamma^\infty_\mathrm{COSMO}$  for the pretrained model in the three validation datasets Val$_\mathrm{ext}$, Val$_\mathrm{edge}$, and Val$_\mathrm{int}$. Mean squared error (MSA) and mean average error (MAE) are shown in the top left corner of every diagram.}
    \label{fig:pre}
\end{figure}

\subsection{Pretraining} \label{sec:pre}

The pretraining of the model on the COSMO dataset takes \SI{34}{\hour} on an RTX~2080~Ti. The resulting model predictions of the three validation sets are shown in a heatmap in Figure~\ref{fig:pre}. For interpolation (Val$_\mathrm{int}$), the pretrained model achieves high accuracy with a mean-squared-error of $\mathrm{MSE} = 0.01$  and a mean-absolute-error of $\mathrm{MAE} = 0.06$. For edge extrapolation (Val$_\mathrm{edge}$), the pretrained model has an MSE of 0.13 and MAE of 0.15, and for extrapolation (Val$_\mathrm{ext}$), an MSE of 0.2 and an MAE of 0.18. The progression of validation and training loss during the pretraining is available in the Supporting Information S5. 

Overall, the resulting error of our pretrained model towards the COSMO-RS data is smaller than the error of COSMO-RS towards experimental data, as reported by \cite{Brouwer.2019}. The result highlights the high interpolation and extrapolation capabilities of our pretrained model for predicting temperature-dependent limiting activity coefficients. Furthermore, the machine learning model is very fast, predicting around 3000 limiting activity coefficients per second on an RTX~2080~Ti without requiring precalculation of sigma surfaces. This high speed should remove property prediction as a bottleneck and allow for the exploration of larger spaces when searching for new components.

\subsection{Fine-Tuning} \label{sec:fine}
\begin{figure}[b!]
    \centering
    \captionsetup[subfigure]{oneside,margin={1cm,0cm}}
    \subfloat[Pretrained]{
    \includegraphics[width=0.45\textwidth]{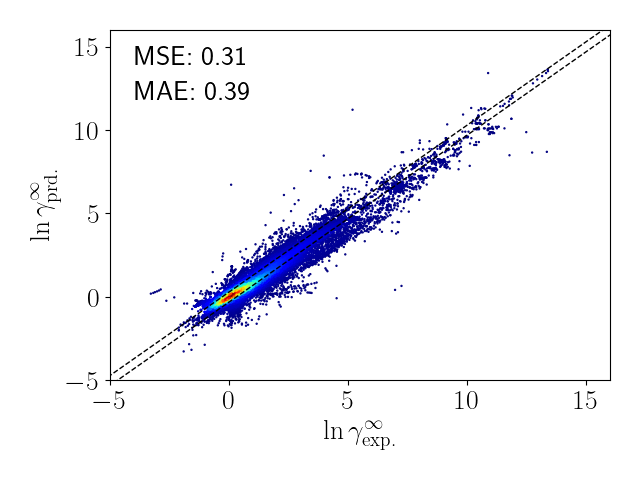}%
    \label{fig:finepre}
    }
    \captionsetup[subfigure]{oneside,margin={1cm,0cm}}
    \subfloat[Val$_\mathrm{int}$]{
    \includegraphics[width=0.45\textwidth]{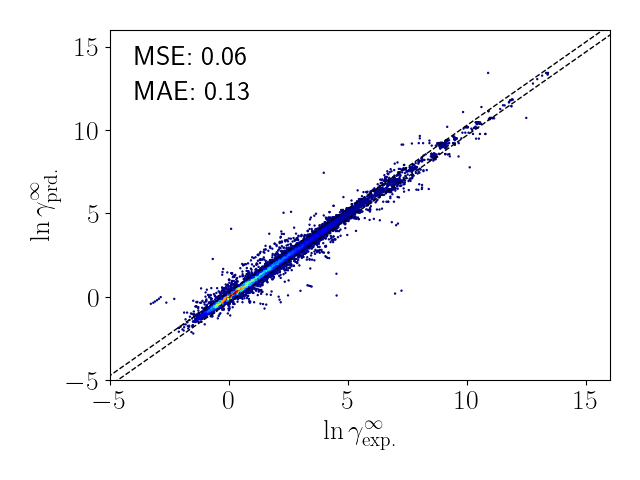}%
    \label{fig:fineint}
    }
    \captionsetup[subfigure]{oneside,margin={1cm,0cm}}
    \subfloat[Val$_\mathrm{edge}$]{
    \includegraphics[width=0.45\textwidth]{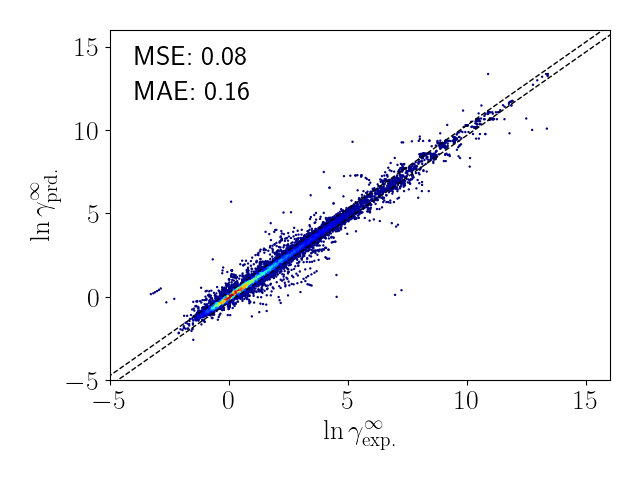}%
    \label{fig:fineedge}
    }
    \captionsetup[subfigure]{oneside,margin={1cm,0cm}}
    \subfloat[Val$_\mathrm{ext}$]{
    \includegraphics[width=0.45\textwidth]{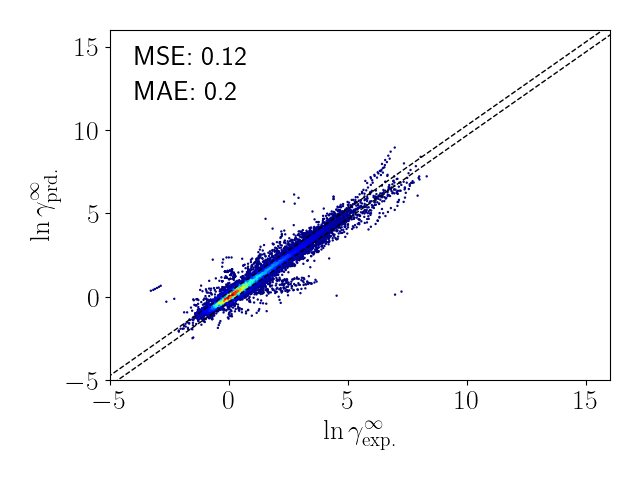}%
    \label{fig:fineext}
    }
    \caption{Predicted vs. experimental limiting activity coefficients from the pretrained model (a), and for the fine-tuned models, Val$_\mathrm{int}$ (b),Val$_\mathrm{edge}$ (c) and Val$_\mathrm{ext}$ (d). For the fine-tuned model, multiple instances of the same molecule can occur in different iterations of Val$_{\mathrm{int},i}$ and Val$_{\mathrm{edge},i}$. For this case the mean of all predictions is shown.}
    \label{fig:fine}
\end{figure}

The fine-tuning was performed on an RTX~2080~Ti and took \SI{6}{\minute} for an individual dataset and \SI{100}{\hour} for all $1\,000$ datasets. The high speed of fine-tuning one dataset enables fine-tuning with single datasets even without a GPU. Fine-tuning on a CPU is expected to be around 200 times slower, thus taking about \SI{20}{\hour} to fine-tune one dataset. 

To analyze the performance of the fine-tuned SPT model, the Brouwer dataset is first predicted using the pretrained model (Figure~\ref{fig:finepre}). The pretrained model achieves an MSE of 0.32 and MAE of 0.39, which is comparable to the accuracy of COSMO-RS for the same dataset (MSE 0.36, MAE 0.38). 

The results of the n-fold cross-validation of the fine-tuned SPT are shown in Figure~\ref{fig:fine}. For interpolation (Val$_{\mathrm{int}}$), fine-tuned SPT archives an MSE of 0.06 and an MAE of 0.13 (Figure~\ref{fig:fineint}) and for edge extrapolation (Val$_{\mathrm{edge}}$), an MSE of 0.08 and an MAE of 0.16 (Figure~\ref{fig:fineedge}). Thus, the prediction of the fine-tuned SPT model for interpolation (Val$_{\mathrm{int}}$) and edge extrapolation (Val$_\mathrm{edge}$) is close to experimental accuracy of between 0.1 and 0.2 \citep{Damay.2021}. However, this high accuracy is only achieved if at least one of the mixture components is included in the training set. Still, for the extrapolation (Val$_\mathrm{ext}$), the MSE and MAE are only slightly higher with 0.12 and 0.20, respectively (Figure~\ref{fig:fineext}). Notably, even in Val$_\mathrm{ext}$, SPT is outperforming COSMO-RS (MSE$_\mathrm{SPT}$ 0.12 vs. MSE$_\mathrm{COSMO\text{-}RS}$ 0.36) (see Section 4). 

The highest errors are mainly obtained for mixture compounds containing nitrogen and silicon. However, only a few data points are contained in the training data with molecules containing silicon. Thus, the prediction might improve with more training data. Overall, the fine-tuning improves the already high accuracy of the pretrained model for all validation sets, leading to a highly accurate prediction of temperature-dependent limiting activity coefficients. Some artifacts seen in Figure~\ref{fig:fine} might also be the result of faulty measurements, as they come from few publications. More curated training and validation data thus might still improve prediction. The results highlight the advantages of combining synthetic and experimental data for predicting thermodynamic properties using deep learning.

\section{Comparison to other models} \label{sec:comp}

To assess the performance of the SPT model discussed in Section\ref{sec:res}, we benchmark our model against competing models from the literature. We first compare our model on temperature-dependent data with the predictive physical models COSMO-RS, UNIFAC, and the recent machine learning approach based on matrix completion by \cite{Damay.2021} (Section~\ref{sec:brouwer}). A comparison to COSMO-SAC implementations is available in the Supporting Information S7. Subsequently, we compare the inter-and extrapolation capabilities of SPT to the graph neural network by \cite{SanchezMedina.2022} on an isothermal dataset by \citet{SanchezMedina.2022} (Section~\ref{sec:medina}). Following \cite{Damay.2021}, we use the percentage of data points with $|\Delta \ln \gamma^\infty| < 0.3 $ as our primary quality measure for the comparison. The percentage of data with $|\Delta \ln \gamma^\infty| < 0.3 $ as well as the mean average error (MAE) and mean squared error (MSE) are summarized in Table \ref{tab:comp}.

\begin{figure}[tb]
\centering
    \subfloat[Brouwer dataset]{
    \includegraphics[width=0.45\textwidth]{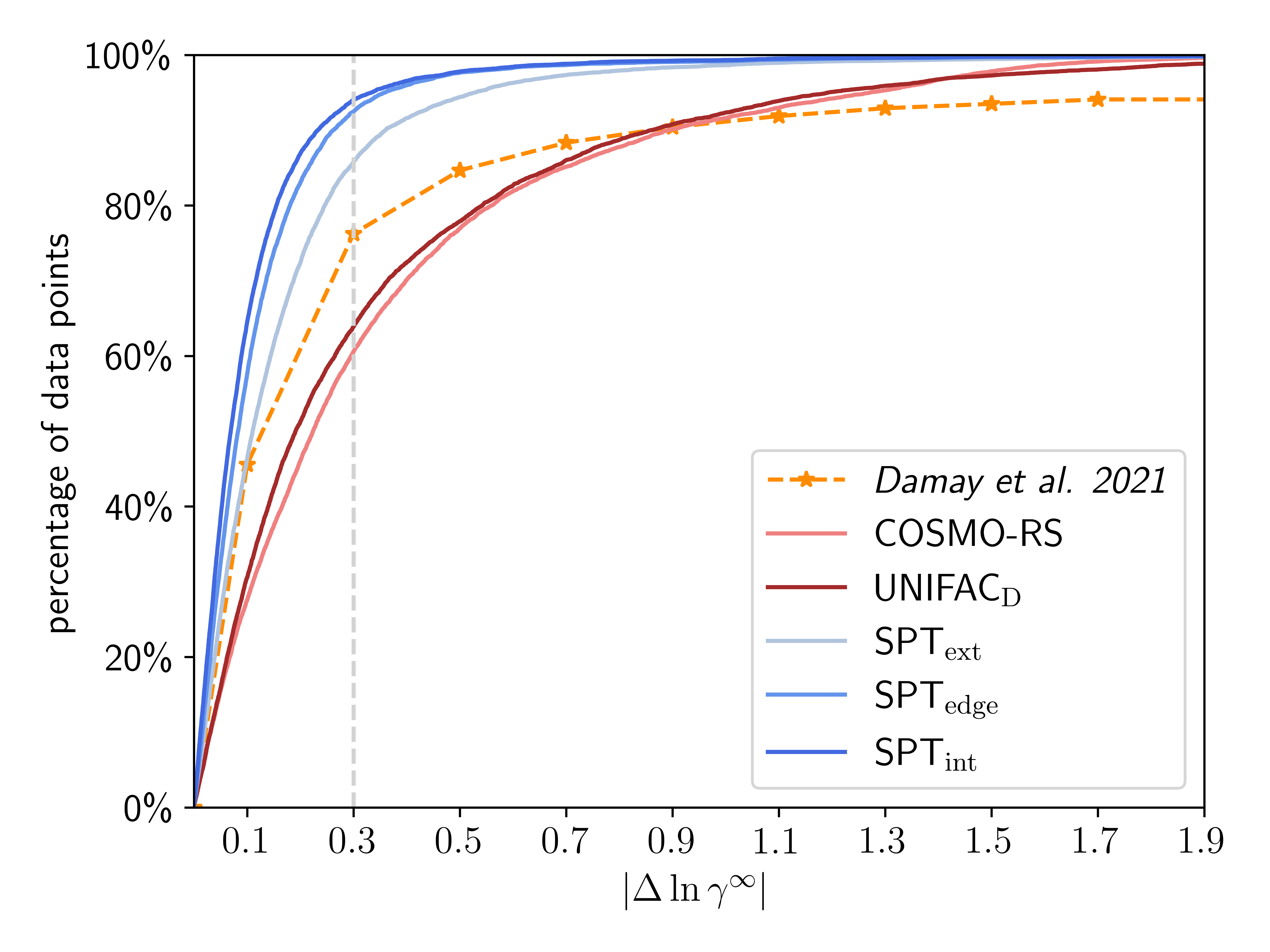}%
    \label{fig:brower}
    }
    \subfloat[Medina dataset]{
    \includegraphics[width=0.45\textwidth]{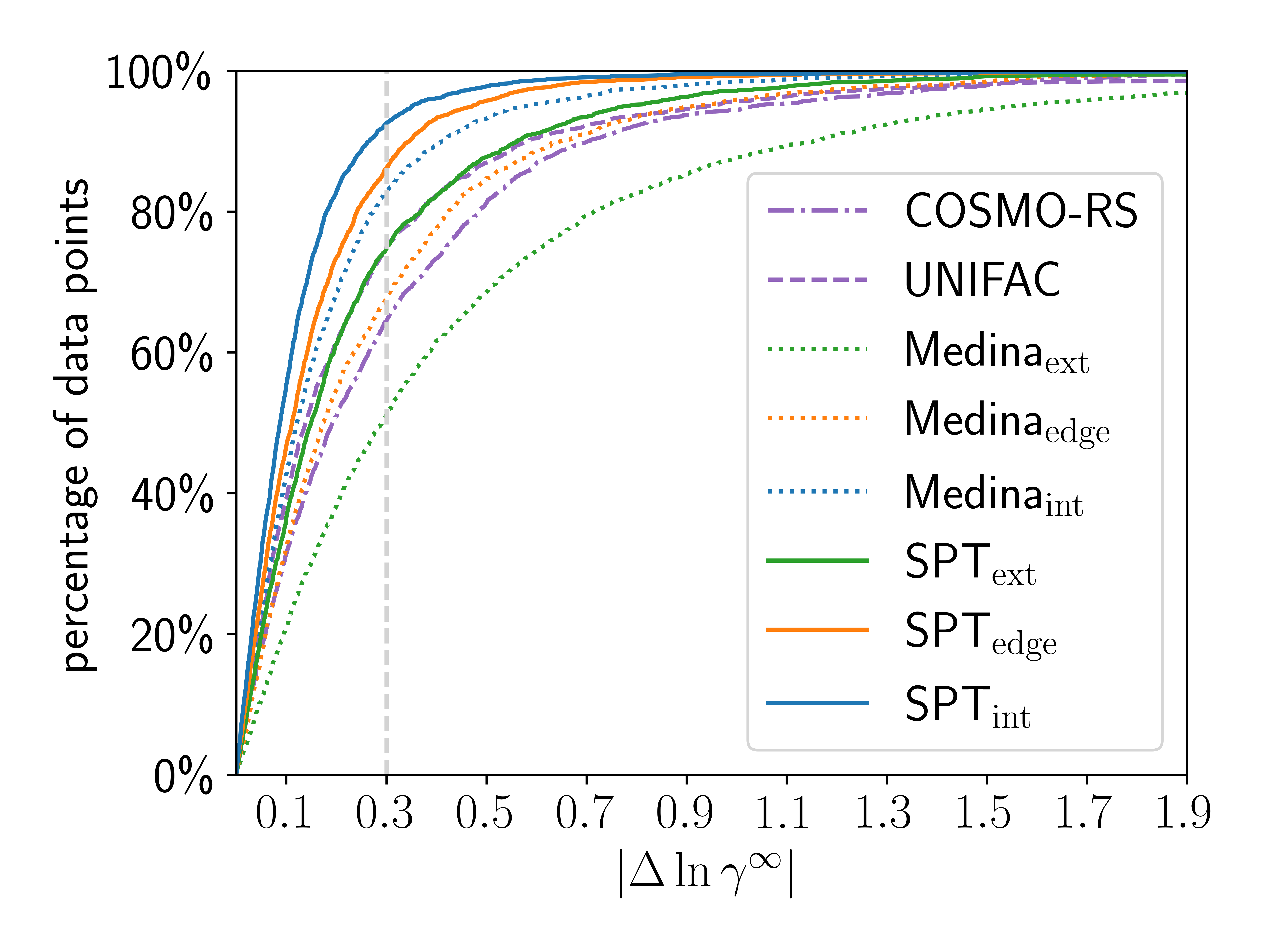}%
    \label{fig:medina}
    }
    \caption{Cumulative distribution of the prediction error for COSMO-RS, UNIFAC, SPT$_\mathrm{ext}$, SPT$_\mathrm{edge}$, SPT$_\mathrm{int}$, Medina$_\mathrm{ext}$, Medina$_\mathrm{edge}$, and Medina$_\mathrm{int}$ using a common subset of the (a) Brower or (b) Medina dataset. For Val$_\mathrm{edge}$ and Val$_\mathrm{int}$ the mean of the n-fold cross-validation is used. Data for \citet{Damay.2021} is approximated from their publication and was evaluated on a different dataset.}
\end{figure}

\begin{table}[b!]
\centering
\caption{Mean average error (MAE), mean square error (MSE), and the percentage of data with $|\Delta \ln \gamma^\infty| < 0.3 $  of the assessed models COSMO-RS, UNIFAC, \cite{Damay.2021}, \citet{SanchezMedina.2022}, and SPT on the common Brouwer and Medina datasets. For performance on all data points see SI9. Generally all models perform slightly worse when considering all datapoints with UNIFAC performing significantly worse. The model of \cite{Damay.2021} does not include MAE and MSE as they are not disclosed in the original publication, and the model is not available for reproduction.}
\begin{tabular}{lcccccc}\toprule
\textbf{Dataset} &\multicolumn{3}{c}{Brouwer} &\multicolumn{3}{c}{Medina}\\ \midrule
\textbf{Error}   &MAE &MSE &$|\Delta \ln \gamma^\infty| < 0.3 $&MAE &MSE &$|\Delta \ln \gamma^\infty| < 0.3 $ \\ \midrule
COSMO-RS &0.36 & 0.29 & 60.6\% &0.31 &0.23 &64.5\% \\
UNIFAC &0.35 &0.45 &63.9\% &0.28 &0.33 &74.9\% \\
Damay et al. (on DDB) &- &- &(\textit{76.6}\%) & & & \\ \midrule
Medina$_\mathrm{ext}$ & & & &0.47 &0.52 &51.1\% \\
Medina$_\mathrm{edge}$ & & &&0.28 &0.20 &67.7\% \\
Medina$_\mathrm{int}$ & & &&0.19 &0.10 &82.8\% \\ \midrule
SPT$_\mathrm{ext}$&\textbf{0.17} &\textbf{0.09} &\textbf{85.8}\% &\textbf{0.25} &\textbf{0.17} &\textbf{74.7}\% \\
SPT$_\mathrm{edge}$&\textbf{0.13} &\textbf{0.06} &\textbf{92.5}\% &\textbf{0.16} &\textbf{0.07} &\textbf{86.1}\% \\
SPT$_\mathrm{int}$ &\textbf{0.11} &\textbf{0.05} &\textbf{94.0}\%&\textbf{0.13} &\textbf{0.05} &\textbf{92.5}\% \\
\bottomrule
\end{tabular}
\label{tab:comp}
\end{table}

\subsection{Comparison on Brouwer dataset} \label{sec:brouwer}

For the comparison on the Brouwer dataset, we calculate all solute/solvent combinations of the Brouwer dataset available in COSMO-RS using the COSMO-RS database 2020 with TZVDP-Fine parametrization and up to 3 conformers. For UNIFAC, we used the UNIFAC$_\mathrm{Dortmund}$ implementation by \cite{Bell.20162022} with 2019 parameters and UNIFAC groups by \cite{Muller.2019}. For a consistent comparison, the results show only the 9625 combinations available in all compared sets, i.e., COSMO-RS database, UNIFAC, and Val$_\mathrm{int}$, Val$_\mathrm{edge}$, and Val$_\mathrm{ext}$ (Figure~\ref{fig:brower}). For Val$_\mathrm{int}$ and Val$_\mathrm{edge}$  the mean of the n-fold validation is used for each mixture. 

The physical models, UNIFAC and COMSO-RS, have very similar performance, with UNIFAC surpassing COSMO-RS slightly with 63.9\% of data below an error of 0.3 compared to 60.5\% for COSMO-RS on the common dataset. COSMO-SAC-based models perform substantially worse than COSMO-RS and UNIFAC (38\% for COSMO-SAC$_\mathrm{2002}$, and 50\% for COSMO-SAC$_\mathrm{dsp}$ see S7). SPT achieves higher accuracy than COSMO-RS and UNIFAC, even for extrapolation Val$_\mathrm{ext}$: Val$_\mathrm{ext}$ predicts 85.8\% of all data points with $|\Delta \ln \gamma^\infty| < 0.3 $ for the compared mixtures. The validation sets Val$_\mathrm{int}$ and Val$_\mathrm{edge}$ achieve even higher accuracies with $|\Delta \ln \gamma^\infty| < 0.3 $ for 94.0\% and 92.5 \% of all combinations, respectively. While our ML model relies on the COSMO models to generate initial data for pretraining, the fine-tuning step on experimental data allows it to surpass the accuracy of the original COSMO models.

In a further analysis, we compare SPT to the machine learning-based model from \cite{Damay.2021}. The authors use matrix completion and train the model to predict limiting activity coefficients from the commercial database DDB. The resulting model yields higher accuracy than the reference model UNIFAC for data taken from the DDB. The authors report that 76.6\% of all data points are within $|\Delta \ln \gamma^\infty| < 0.3 $  when using leave-one-out validation. For qualitative comparison, the results of  Damay et al. (Figure 10 of \cite{Damay.2021}) are shown in Figure \ref{fig:scaling}. This result is most comparable to our validation set Val$_\mathrm{int}$ (94.0 \% with $|\Delta \ln \gamma^\infty| < 0.3 $) since matrix completion only allows interpolation when both molecules are contained within the training. However, since the authors used another non-public dataset for training (DDB), these results are not directly comparable to our results. Comparing UNIFAC to both datasets, Damay et al. report a higher accuracy of UNIFAC on the DDB dataset than we obtain for UNIFAC on the Brouwer dataset (71\% with $|\Delta \ln \gamma^\infty| < 0.3 $ for UNIFAC on DDB vs. 63\% for Brouwer). This result can indicate that the data in the DDB is of better quality. Thus, SPT's performance may improve in performance when fine-tuned on the DDB data. 

In contrast to matrix completion, SPT allows for extrapolating unseen and partly unseen solute/solvent combinations. The (edge) extrapolation capacity of our model indicates a high accuracy even if compared to the interpolation accuracy of the matrix completion model proposed by \cite{Damay.2021} with 85.8\% and 92.5\% of all data points with $|\Delta \ln \gamma^\infty| < 0.3 $, respectively. While evaluation took place on different datasets and thus results are not directly comparable, these results still strongly suggest that SPT can achieve higher accuracies in predicting limiting activity coefficients than matrix completion, though coming at a higher computational effort.

\subsection{Comparison on Medina dataset} \label{sec:medina}

\cite{SanchezMedina.2022} proposed a graph neural network for predicting limiting activity coefficients at \SI{298.15}{\kelvin}. An extension for temperature dependency is proposed in the outlook but not yet available in the model. The authors tested the model using random splits, resulting in sets most comparable to our Val$_\mathrm{int}$ set. Thus, the extrapolation capabilities of the model proposed by \citet{SanchezMedina.2022} are unknown.

For a consistent comparison of the SPT model and the Medina model, we split the dataset from \cite{SanchezMedina.2022} (Medina dataset) into 200 training and validation sets according to our validation strategy discussed in Section~\ref{sec:data_exp}. Subsequently, we train the Medina model and our model on the resulting 200 training sets (SI S7). Due to the lack of a test set to stop training and adjust the learning rate, we use the performance on Val$_\mathrm{ext}$ to set the learning rate and select the epoch with the lowest mean validation MSE out of the 200 training epochs across the 200 datasets for each validation set (Val$_\mathrm{ext}$~=~\num{117}, Val$_\mathrm{edge}$~=~\num{135}, Val$_\mathrm{int}$~=~\num{163}). For SPT, we use the performance at the final epoch (50) as previously. As in Section~\ref{sec:brouwer}, for Val$_\mathrm{int}$ and Val$_\mathrm{edge}$, the mean of the n-fold validation is calculated and used for each unique mixture. For the Medina model training failed on the sets 87, 115, 149 and 182 for unknown reasons, these sets are excluded.

The MSE and MAE of the Medina model on Val$_\mathrm{int}$ as calculated by us (MSE: \num{0.10}, MAE: \num{0.19}) reproduces the MSE and MAE reported by \cite{SanchezMedina.2022} using random splitting (MSE: 0.10, MAE: 0.18) (Table \ref{tab:comp}). This result indicates that random splitting results in a test set that is similar to our Val$_\mathrm{int}$ set and random splitting is thus not suitable to assess the extrapolation capabilities of models.

Figure \ref{fig:medina} shows the prediction error of COSMO-RS, UNIFAC$_\mathrm{Dortmund}$, the Medina model, and the SPT model fine-tuned on the Medina dataset. The Medina dataset is reduced from 2810 mixtures to the 2469 mixtures that all models can calculate. 

SPT  generally outperforms the Medina model on all validation sets. For Val$_\mathrm{int}$, 92.5\% of the data points are with $|\Delta \ln \gamma^\infty| < 0.3 $ for SPT compared to 82.8\% for the Medina model. For Val$_\mathrm{edge}$, 86.1\% and 67.7\%, and for Val$_\mathrm{ext}$, 74.9\% and 51.1\% of data data points are with $|\Delta \ln \gamma^\infty| < 0.3 $ for SPT and the Medina model, respectively. The MAE of SPT is about half of the MAE of the Medina model for each validation set. Particularly, the vast difference in performance for (edge) extrapolation highlights the effective performance of SPT when predicting new molecules.
As for the Brouwer dataset (Section~\ref{sec:brouwer}), SPT outperforms COSMO-RS and UNIFAC on the Medina dataset even for extrapolation. Similarly, the Medina model outperforms COSMO-RS and UNIFAC for interpolation tasks, but performs similar to COSMO-RS and worse than UNIFAC on edge extrapolation and is surpassed for extrapolation by both COSMO-RS and UNIFAC. Please note that it is very likely that UNIFAC parameters were fitted to mixtures contained in the Medina dataset, likely improving the UNIFAC performance for this dataset. 

The results highlight the advantage of our pretraining on synthetic data to exploit scarce experimental data and extend the extrapolative abilities of our model. The obtained data-driven model shows a good understanding of molecular properties. Overall, SPT performs slightly worse on the Medina dataset than on the Brouwer dataset, likely due to the smaller total amount of training data (2 810 vs.  $20\,870$). Therefore, we analyze the data scaling of our SPT model in more detail in Section~\ref{sec:scaling}.

Additionally to the increased accuracy, our SPT model requires \SI{45}{\second} to fine-tune for 50 epochs on the Medina dataset, while the Medina model requires around \SI{4}{\minute} for 50 epochs on an RTX~2080~Ti, even though the Medina model has much fewer parameters ($21\,000$ vs. 6.5 million). The shorter training time can be vital if no GPU is available. However, the training time of the Medina model would likely be improved with the use of mixed-precision training, and SPT requires lengthy pretraining before fine-tuning.

\section{Data scaling of the model} \label{sec:scaling}

In Section~\ref{sec:res}, the SPT model was trained using on average $17\,370$ data points from the Brouwer dataset. Machine learning models are well known for increasing their performance with larger amounts of training data. Conversely, for many thermodynamic properties, less experimental data is available than for limiting activity coefficients. Thus, this section gives insight into SPT's data scaling to estimate model improvements with larger datasets and the expected model performance when less experimental data is available for fine-tuning.

To determine the scaling of the fine-tuning of the SPT model, we create 200 training datasets, each containing n$_\mathrm{train}$ random unique solute/solvent combinations from the Brouwer dataset for n$_\mathrm{train}$ between 2 and 5000 solute/solvent combinations excluding water. The remaining solute/solvent combinations in the Brouwer dataset are then sorted into the validation sets Val$_\mathrm{ext}$, Val$_\mathrm{edge}$,  and Val$_\mathrm{int}$. For large numbers n$_\mathrm{train}$, only a few solute/solvent combinations remain in Val$_\mathrm{ext}$, and Val$_\mathrm{edge}$, since common molecules are likely to be included in the training dataset and thus necessarily excluded from the validation sets Val$_\mathrm{ext}$, and Val$_\mathrm{edge}$. For example, for 5000 training mixtures, only 17 unique solute/solvent combinations remain in the validation set Val$_\mathrm{ext}$,5000 across all 200 training datasets. Moreover, many of the 200 training datasets do not have a single solute/solvent combination in the validation set Val$_\mathrm{ext}$. This small number of solute/solvent combinations for validation set Val$_\mathrm{ext}$ leads to high variance. Thus, we only consider validation sets Val$_\mathrm{ext}$ and Val$_\mathrm{edge}$ where more than \num{17500} of the \num{18348} solute/solvent combinations are still present. The cutoff point is n$_\mathrm{train}$ = \num{80} for Val$_\mathrm{ext}$ and n$_\mathrm{train}$ = \num{500} for Val$_\mathrm{edge}$. For Val$_\mathrm{int}$ the reverse is the case. Here, small n$_\mathrm{train}$ lead to unreliable results and thus, no n$_\mathrm{train}$ below \num{50} is considered. 

\begin{figure}[tb]
    \centering
    \includegraphics[width=0.5\textwidth]{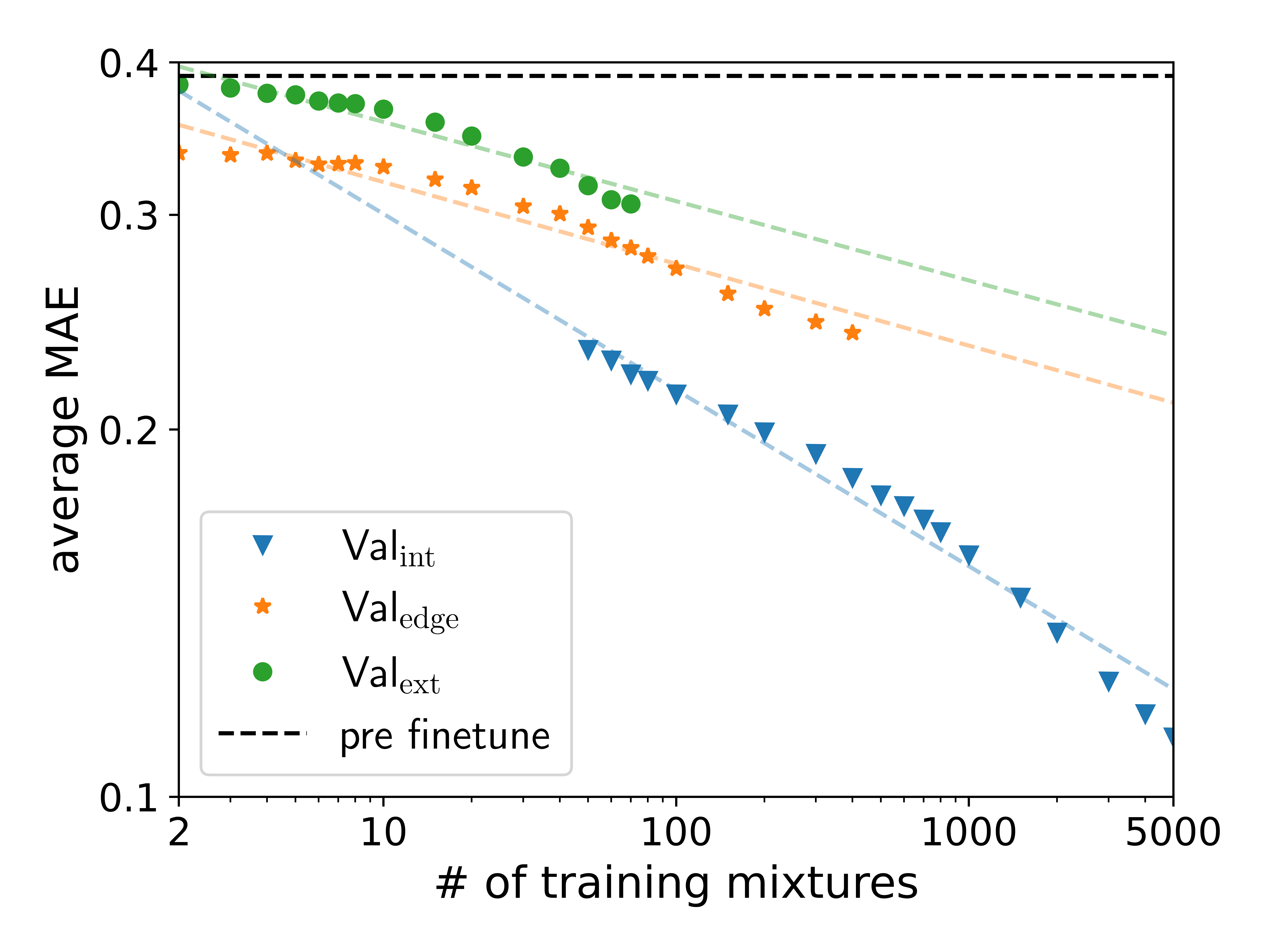}
    \caption{Scaling behavior of SPT's average MAE for the data sets Val$_\mathrm{ext}$, Val$_\mathrm{edge}$ and Val$_\mathrm{int}$ as function of available experimental data for fine-tuning. The solid line indicates the performance of the pretrained model without fine-tuning}
    \label{fig:scaling}
\end{figure}

The MAE of Val$_\mathrm{ext}$ and Val$_\mathrm{edge}$ decreases linearly with the size of the training dataset in the log-log space (Figure \ref{fig:scaling}). The MAE of Val$_\mathrm{int}$ decreases with a steeper slope, indicating that interpolation might be easier to learn. Furthermore, there is some indication the slope is increasing for even larger training sets. For the investigated training sizes, no saturation is visible in any validation set, indicating that the accuracy of the machine learning model still improves for increasing amounts of experimental data for fine-tuning. Following this prediction, between $10\,000$ and 20 000 solute/solvent combinations would be needed for training to reach an average MAE of lower than 0.15 for Val$_\mathrm{ext}$ , which is within experimental accuracy. The amount of required data would thus be smaller than the $31\,000$ unique solute/solvent combinations available in the commercial database DDB, indicating that high-quality prediction of limiting activity coefficients is in reach.

Even small amounts of experimental data used for fine-tuning lead to substantial improvements in the validation set Val$_\mathrm{edge}$. SPT should thus 
require only a few experimental data points for fine-tuning for accurate predictions around specific data points. The high performance of SPT, even for limited experimental data available, originates from the pretraining to synthetic data, which enables learning the underlying grammar of the molecular representation and the physics provided by the predictive thermodynamic model used to generate the synthetic data. The capability of our model to accurately predict similar mixtures with only a few experimental data points could be used to guide experiments by measuring and predicting in tandem, narrowing down a target region. 

\section{Conclusions} \label{sec:con}

One of the main roadblocks to the widespread application of deep learning in chemical engineering is the availability of training data. Particularly, for predicting thermodynamic mixture properties, often only a limited amount of experimental data is available. This work tackles the challenge of scarce data availability for thermodynamic property prediction based on deep learning by combining synthetic data with experimental data. For this purpose, we introduce the SPT model, which we pretrain to synthetic data generated using COSMO-RS and subsequently fine-tune the model using experimental data. Thereby, we achieve a highly accurate prediction of temperature-dependent limiting activity coefficients solely from SMILES codes. 

The SPT machine learning model surpasses the accuracy of conventional predictive thermodynamic models such as COSMO-SAC, COSMO-RS, and UNIFAC and recently proposed machine learning approaches based on matrix completion and graph neural networks.

Combining synthetic data with scarce experimental data opens new possibilities for the training of deep learning models for thermodynamic property prediction. Even small amounts of experimental data points already lead to significant improvements in the prediction quality of SPT. Furthermore, the main computational effort is in the pretraining of the model to synthetic data, while the fine-tuning is computationally efficient. 

The efficient fine-tuning opens up possibilities to combine deep learning with automated experiments, where a model is continuously refined with experimental data while providing predictions of new promising candidates to measure. Such workflows could generate machine learning models that are highly accurate in specific domains. 

\section{Available data}
\begin{itemize}
    \item \href{https://github.com/Bene94/SMILES2PropertiesTransformer}{https://github.com/Bene94/SMILES2PropertiesTransformer}
    \item Datasets and trained models \href{https://polybox.ethz.ch/index.php/s/kyVOt3pwHW26PP4}{www.polybox.ethz.ch/}
\end{itemize}

\bibliographystyle{abbrvnat}
\bibliography{lit}  

\begin{thebibliography}{39}
\providecommand{\natexlab}[1]{#1}
\providecommand{\url}[1]{\texttt{#1}}
\expandafter\ifx\csname urlstyle\endcsname\relax
  \providecommand{\doi}[1]{doi: #1}\else
  \providecommand{\doi}{doi: \begingroup \urlstyle{rm}\Url}\fi

\bibitem[Alammar(2018)]{Alammar.2018}
J.~Alammar.
\newblock {The Illustrated Transformer}, 2018.
\newblock URL \url{https://jalammar.github.io/illustrated-transformer/}.

\bibitem[Alshehri and You(2022)]{Alshehri.2022}
A.~S. Alshehri and F.~You.
\newblock {Machine learning for multiscale modeling in computational molecular
  design}.
\newblock \emph{{Current Opinion in Chemical Engineering}}, 36:\penalty0
  100752, 2022.
\newblock ISSN 22113398.
\newblock \doi{10.1016/j.coche.2021.100752}.

\bibitem[Alshehri et~al.(2021)Alshehri, Tula, You, and Gani]{Alshehri.2021}
A.~S. Alshehri, A.~K. Tula, F.~You, and R.~Gani.
\newblock {Next generation pure component property estimation models: With and
  without machine learning techniques}.
\newblock \emph{{AIChE Journal}}, 2021.
\newblock ISSN 0001-1541.
\newblock \doi{10.1002/aic.17469}.

\bibitem[Bell and Contributers(2016-2022)]{Bell.20162022}
C.~Bell and Contributers.
\newblock {Thermo: Chemical properties component of Chemical Engineering Design
  Library (ChEDL)}, 2016-2022.
\newblock URL \url{https://github.com/CalebBell/thermo.}

\bibitem[Bjerrum(2017)]{Bjerrum.21.03.2017}
E.~J. Bjerrum.
\newblock {SMILES Enumeration as Data Augmentation for Neural Network Modeling
  of Molecules}, 2017.
\newblock URL \url{https://arxiv.org/pdf/1703.07076}.

\bibitem[Brouwer and Schuur(2019)]{Brouwer.2019}
T.~Brouwer and B.~Schuur.
\newblock {Model Performances Evaluated for Infinite Dilution Activity
  Coefficients Prediction at 298.15 K}.
\newblock \emph{{Industrial {\&} Engineering Chemistry Research}}, 58\penalty0
  (20):\penalty0 8903--8914, 2019.
\newblock ISSN 0888-5885.
\newblock \doi{10.1021/acs.iecr.9b00727}.

\bibitem[Brouwer et~al.(2021)Brouwer, Kersten, Bargeman, and
  Schuur]{Brouwer.2021}
T.~Brouwer, S.~R. Kersten, G.~Bargeman, and B.~Schuur.
\newblock trends in solvent impact on infinite dilution activity coefficients
  of solutes reviewed and visualized using an algorithm to support selection of
  solvents for greener fluid separations.
\newblock \emph{{Separation and Purification Technology}}, 272:\penalty0
  118727, 2021.
\newblock ISSN 13835866.
\newblock \doi{10.1016/j.seppur.2021.118727}.

\bibitem[Brown et~al.(2020)Brown, Mann, Ryder, Subbiah, Kaplan, Dhariwal,
  Neelakantan, Shyam, Sastry, Askell, Agarwal, Herbert-Voss, Krueger, Henighan,
  Child, Ramesh, Ziegler, Wu, Winter, Hesse, Chen, Sigler, Litwin, Gray, Chess,
  Clark, Berner, McCandlish, Radford, Sutskever, and Amodei]{Brown.28.05.2020}
T.~B. Brown, B.~Mann, N.~Ryder, M.~Subbiah, J.~Kaplan, P.~Dhariwal,
  A.~Neelakantan, P.~Shyam, G.~Sastry, A.~Askell, S.~Agarwal, A.~Herbert-Voss,
  G.~Krueger, T.~Henighan, R.~Child, A.~Ramesh, D.~M. Ziegler, J.~Wu,
  C.~Winter, C.~Hesse, M.~Chen, E.~Sigler, M.~Litwin, S.~Gray, B.~Chess,
  J.~Clark, C.~Berner, S.~McCandlish, A.~Radford, I.~Sutskever, and D.~Amodei.
\newblock {Language Models are Few-Shot Learners}, 2020.
\newblock URL \url{http://arxiv.org/pdf/2005.14165v4}.

\bibitem[CAS(2022)]{CAS.06.02.2022}
CAS, 2022.
\newblock URL \url{https://commonchemistry.cas.org/}.

\bibitem[Chen et~al.(2021{\natexlab{a}})Chen, Song, and Qi]{Chen.2021}
G.~Chen, Z.~Song, and Z.~Qi.
\newblock {Transformer-convolutional neural network for surface charge density
  profile prediction: Enabling high-throughput solvent screening with
  COSMO-SAC}.
\newblock \emph{{Chemical Engineering Science}}, 246:\penalty0 117002,
  2021{\natexlab{a}}.
\newblock ISSN 00092509.
\newblock \doi{10.1016/j.ces.2021.117002}.

\bibitem[Chen et~al.(2021{\natexlab{b}})Chen, Song, Qi, and
  Sundmacher]{Chen.2021b}
G.~Chen, Z.~Song, Z.~Qi, and K.~Sundmacher.
\newblock {Neural recommender system for the activity coefficient prediction
  and UNIFAC model extension of ionic liquid--solute systems}.
\newblock \emph{{AIChE Journal}}, 67\penalty0 (4), 2021{\natexlab{b}}.
\newblock ISSN 0001-1541.
\newblock \doi{10.1002/aic.17171}.

\bibitem[Damay et~al.(2021)Damay, Jirasek, Kloft, Bortz, and Hasse]{Damay.2021}
J.~Damay, F.~Jirasek, M.~Kloft, M.~Bortz, and H.~Hasse.
\newblock {Predicting Activity Coefficients at Infinite Dilution for Varying
  Temperatures by Matrix Completion}.
\newblock \emph{{Industrial {\&} Engineering Chemistry Research}}, 60\penalty0
  (40):\penalty0 14564--14578, 2021.
\newblock ISSN 0888-5885.
\newblock \doi{10.1021/acs.iecr.1c02039}.

\bibitem[Dobbelaere et~al.(2021)Dobbelaere, Plehiers, {van de Vijver}, Stevens,
  and {van Geem}]{Dobbelaere.2021}
M.~R. Dobbelaere, P.~P. Plehiers, R.~{van de Vijver}, C.~V. Stevens, and K.~M.
  {van Geem}.
\newblock {Machine Learning in Chemical Engineering: Strengths, Weaknesses,
  Opportunities, and Threats}.
\newblock \emph{{Engineering}}, 7\penalty0 (9):\penalty0 1201--1211, 2021.
\newblock ISSN 20958099.
\newblock \doi{10.1016/j.eng.2021.03.019}.

\bibitem[{Dortmund Datenbank}(2022)]{DortmundDatenbank.2022}
{Dortmund Datenbank}, 2022.
\newblock URL \url{http://www.ddbst.com/}.

\bibitem[Dosovitskiy et~al.(2020)Dosovitskiy, Beyer, Kolesnikov, Weissenborn,
  Zhai, Unterthiner, Dehghani, Minderer, Heigold, Gelly, Uszkoreit, and
  Houlsby]{Dosovitskiy.22.10.2020}
A.~Dosovitskiy, L.~Beyer, A.~Kolesnikov, D.~Weissenborn, X.~Zhai,
  T.~Unterthiner, M.~Dehghani, M.~Minderer, G.~Heigold, S.~Gelly, J.~Uszkoreit,
  and N.~Houlsby.
\newblock {An Image is Worth 16x16 Words: Transformers for Image Recognition at
  Scale}, 2020.
\newblock URL \url{http://arxiv.org/pdf/2010.11929v2}.

\bibitem[Fredenslund et~al.(1975)Fredenslund, Jones, and
  Prausnitz]{Fredenslund.1975}
A.~Fredenslund, R.~L. Jones, and J.~M. Prausnitz.
\newblock {Group-contribution estimation of activity coefficients in nonideal
  liquid mixtures}.
\newblock \emph{{AIChE Journal}}, 21\penalty0 (6):\penalty0 1086--1099, 1975.
\newblock ISSN 0001-1541.
\newblock \doi{10.1002/aic.690210607}.

\bibitem[Haghighatlari and Hachmann(2019)]{Haghighatlari.2019}
M.~Haghighatlari and J.~Hachmann.
\newblock {Advances of machine learning in molecular modeling and simulation}.
\newblock \emph{{Current Opinion in Chemical Engineering}}, 23:\penalty0
  51--57, 2019.
\newblock ISSN 22113398.
\newblock \doi{10.1016/j.coche.2019.02.009}.

\bibitem[Honda et~al.(2019)Honda, Shi, and Ueda]{Honda.12.11.2019}
S.~Honda, S.~Shi, and H.~R. Ueda.
\newblock {SMILES Transformer: Pre-trained Molecular Fingerprint for Low Data
  Drug Discovery}, 2019.
\newblock URL \url{http://arxiv.org/pdf/1911.04738v1}.

\bibitem[Jirasek et~al.(2020)Jirasek, Alves, Damay, Vandermeulen, Bamler,
  Bortz, Mandt, Kloft, and Hasse]{Jirasek.2020}
F.~Jirasek, R.~A.~S. Alves, J.~Damay, R.~A. Vandermeulen, R.~Bamler, M.~Bortz,
  S.~Mandt, M.~Kloft, and H.~Hasse.
\newblock {Machine Learning in Thermodynamics: Prediction of Activity
  Coefficients by Matrix Completion}.
\newblock \emph{{The journal of physical chemistry letters}}, 11\penalty0
  (3):\penalty0 981--985, 2020.
\newblock \doi{10.1021/acs.jpclett.9b03657}.

\bibitem[Karpathy(2021)]{Karpathy.2021}
A.~Karpathy.
\newblock {minGPT}, 2021.
\newblock URL \url{https://github.com/karpathy/minGPT/blob/master/LICENSE}.

\bibitem[Kim et~al.(2021)Kim, Na, and Lee]{Kim.2021}
H.~Kim, J.~Na, and W.~B. Lee.
\newblock {Generative Chemical Transformer: Neural Machine Learning of
  Molecular Geometric Structures from Chemical Language via Attention}.
\newblock \emph{{Journal of chemical information and modeling}}, 61\penalty0
  (12):\penalty0 5804--5814, 2021.
\newblock \doi{10.1021/acs.jcim.1c01289}.

\bibitem[Klamt(1995)]{Klamt.1995}
A.~Klamt.
\newblock {Conductor-like Screening Model for Real Solvents: A New Approach to
  the Quantitative Calculation of Solvation Phenomena}.
\newblock \emph{{The Journal of Physical Chemistry}}, 99\penalty0 (7):\penalty0
  2224--2235, 1995.
\newblock ISSN 0022-3654.
\newblock \doi{10.1021/j100007a062}.

\bibitem[Lafitte et~al.(2013)Lafitte, Apostolakou, Avenda{\~n}o, Galindo,
  Adjiman, M{\"u}ller, and Jackson]{Lafitte.2013}
T.~Lafitte, A.~Apostolakou, C.~Avenda{\~n}o, A.~Galindo, C.~S. Adjiman, E.~A.
  M{\"u}ller, and G.~Jackson.
\newblock {Accurate statistical associating fluid theory for chain molecules
  formed from Mie segments}.
\newblock \emph{{The Journal of chemical physics}}, 139\penalty0 (15):\penalty0
  154504, 2013.
\newblock \doi{10.1063/1.4819786}.

\bibitem[Lim and Lee(2021)]{Lim.2021}
S.~Lim and Y.~O. Lee.
\newblock {Predicting Chemical Properties using Self-Attention Multi-task
  Learning based on SMILES Representation}.
\newblock In \emph{{2020 25th International Conference on Pattern Recognition
  (ICPR)}}, pages 3146--3153. IEEE, 2021.
\newblock ISBN 978-1-7281-8808-9.
\newblock \doi{10.1109/ICPR48806.2021.9412555}.

\bibitem[Lin and Sandler(2002)]{Lin.2002}
S.-T. Lin and S.~I. Sandler.
\newblock {A Priori Phase Equilibrium Prediction from a Segment Contribution
  Solvation Model}.
\newblock \emph{{Industrial {\&} Engineering Chemistry Research}}, 41\penalty0
  (5):\penalty0 899--913, 2002.
\newblock ISSN 0888-5885.
\newblock \doi{10.1021/ie001047w}.

\bibitem[M{\"u}ller(2019)]{Muller.2019}
S.~M{\"u}ller.
\newblock {Flexible heuristic algorithm for automatic molecule fragmentation:
  application to the UNIFAC group contribution model}.
\newblock \emph{{Journal of cheminformatics}}, 11\penalty0 (1):\penalty0 57,
  2019.
\newblock ISSN 1758-2946.
\newblock \doi{10.1186/s13321-019-0382-3}.

\bibitem[Nebig and Gmehling(2010)]{Nebig.2010}
S.~Nebig and J.~Gmehling.
\newblock {Measurements of different thermodynamic properties of systems
  containing ionic liquids and correlation of these properties using modified
  UNIFAC (Dortmund)}.
\newblock \emph{{Fluid Phase Equilibria}}, 294\penalty0 (1-2):\penalty0
  206--212, 2010.
\newblock ISSN 03783812.
\newblock \doi{10.1016/j.fluid.2010.02.010}.

\bibitem[Parmar et~al.(2018)Parmar, Vaswani, Uszkoreit, Kaiser, Shazeer, Ku,
  and Tran]{Parmar.15.02.2018}
N.~Parmar, A.~Vaswani, J.~Uszkoreit, {\L}.~Kaiser, N.~Shazeer, A.~Ku, and
  D.~Tran.
\newblock {Image Transformer}.
\newblock \emph{https://arxiv.org/pdf/1802.05751}, 2018.
\newblock URL \url{https://arxiv.org/pdf/1802.05751}.

\bibitem[Pytorch(2021)]{Pytorch.2021}
Pytorch, 2021.
\newblock URL
  \url{https://pytorch.org/docs/stable/generated/torch.nn.Transformer.html}.

\bibitem[Rong et~al.(2020)Rong, Bian, Xu, Xie, Wei, Huang, and
  Huang]{Rong.18.06.2020}
Y.~Rong, Y.~Bian, T.~Xu, W.~Xie, Y.~Wei, W.~Huang, and J.~Huang.
\newblock {Self-Supervised Graph Transformer on Large-Scale Molecular Data}.
\newblock \emph{http://arxiv.org/pdf/2007.02835v2}, 2020.
\newblock URL \url{http://arxiv.org/pdf/2007.02835v2}.

\bibitem[{Sanchez Medina} et~al.(2022){Sanchez Medina}, Linke, Stoll, and
  Sundmacher]{SanchezMedina.2022}
E.~I. {Sanchez Medina}, S.~Linke, M.~Stoll, and K.~Sundmacher.
\newblock {Graph neural networks for the prediction of infinite dilution
  activity coefficients}.
\newblock \emph{{Digital Discovery}}, 2022.
\newblock \doi{10.1039/D1DD00037C}.

\bibitem[Scheffczyk et~al.(2016)Scheffczyk, Redepenning, Jens, Winter,
  Leonhard, Marquardt, and Bardow]{Scheffczyk.2016}
J.~Scheffczyk, C.~Redepenning, C.~M. Jens, B.~Winter, K.~Leonhard,
  W.~Marquardt, and A.~Bardow.
\newblock {Massive, automated solvent screening for minimum energy demand in
  hybrid extraction--distillation using COSMO-RS}.
\newblock \emph{{Chemical Engineering Research and Design}}, 115:\penalty0
  433--442, 2016.
\newblock ISSN 02638762.
\newblock \doi{10.1016/j.cherd.2016.09.029}.

\bibitem[Schweidtmann et~al.(2021)Schweidtmann, Esche, Fischer, Kloft, Repke,
  Sager, and Mitsos]{Schweidtmann.2021}
A.~M. Schweidtmann, E.~Esche, A.~Fischer, M.~Kloft, J.-U. Repke, S.~Sager, and
  A.~Mitsos.
\newblock {Machine Learning in Chemical Engineering: A Perspective}.
\newblock \emph{{Chemie Ingenieur Technik}}, 93\penalty0 (12):\penalty0
  2029--2039, 2021.
\newblock ISSN 0009-286X.
\newblock \doi{10.1002/cite.202100083}.

\bibitem[Skinnider et~al.(2021)Skinnider, Wang, Pasin, Greiner, Foster,
  Dalsgaard, and Wishart]{Skinnider.2021}
M.~A. Skinnider, F.~Wang, D.~Pasin, R.~Greiner, L.~J. Foster, P.~W. Dalsgaard,
  and D.~S. Wishart.
\newblock {A deep generative model enables automated structure elucidation of
  novel psychoactive substances}.
\newblock \emph{{Nature Machine Intelligence}}, 3\penalty0 (11):\penalty0
  973--984, 2021.
\newblock \doi{10.1038/s42256-021-00407-x}.

\bibitem[Tetko et~al.(2020)Tetko, Karpov, {van Deursen}, and Godin]{Tetko.2020}
I.~V. Tetko, P.~Karpov, R.~{van Deursen}, and G.~Godin.
\newblock {State-of-the-art augmented NLP transformer models for direct and
  single-step retrosynthesis}.
\newblock \emph{{Nature communications}}, 11\penalty0 (1):\penalty0 5575, 2020.
\newblock \doi{10.1038/s41467-020-19266-y}.

\bibitem[Vaswani et~al.(2017)Vaswani, Shazeer, Parmar, Uszkoreit, Jones, Gomez,
  Kaiser, and Polosukhin]{Vaswani.2017}
A.~Vaswani, N.~Shazeer, N.~Parmar, J.~Uszkoreit, L.~Jones, A.~N. Gomez, b.~u.
  Kaiser, and I.~Polosukhin.
\newblock {Attention is All you Need}.
\newblock In I.~Guyon, U.~V. Luxburg, S.~Bengio, H.~Wallach, R.~Fergus,
  S.~Vishwanathan, and R.~Garnett, editors, \emph{{Advances in Neural
  Information Processing Systems}}, volume~30. {Curran Associates, Inc}, 2017.
\newblock URL
  \url{https://proceedings.neurips.cc/paper/2017/file/3f5ee243547dee91fbd053c1c4a845aa-Paper.pdf}.

\bibitem[Wang et~al.(2019)Wang, Guo, Wang, Sun, and Huang]{Wang.2019}
S.~Wang, Y.~Guo, Y.~Wang, H.~Sun, and J.~Huang.
\newblock {SMILES-BERT}.
\newblock In X.~Shi, M.~Buck, J.~Ma, and P.~Veltri, editors, \emph{{Proceedings
  of the 10th ACM International Conference on Bioinformatics, Computational
  Biology and Health Informatics}}, pages 429--436, New York, NY, USA, 2019.
  ACM.
\newblock ISBN 9781450366663.
\newblock \doi{10.1145/3307339.3342186}.

\bibitem[Weininger(1988)]{Weininger.1988}
D.~Weininger.
\newblock {SMILES, a chemical language and information system. 1. Introduction
  to methodology and encoding rules}.
\newblock \emph{{Journal of chemical information and modeling}}, 28\penalty0
  (1):\penalty0 31--36, 1988.
\newblock ISSN 1549-9596.
\newblock \doi{10.1021/ci00057a005}.

\bibitem[Xiong et~al.(2020)Xiong, Yang, {Di He}, Zheng, Zheng, Xing, Zhang,
  Lan, Wang, and Liu]{Xiong.}
R.~Xiong, Y.~Yang, {Di He}, K.~Zheng, S.~Zheng, C.~Xing, H.~Zhang, Y.~Lan,
  L.~Wang, and T.-Y. Liu.
\newblock {On Layer Normalization in the Transformer Architecture}.
\newblock \emph{http://arxiv.org/pdf/2002.04745v2}, 2020.
\newblock URL \url{http://arxiv.org/pdf/2002.04745v2}.

\end{thebibliography}






\end{document}